\documentclass[a4paper,10pt]{article}
\usepackage[utf8]{inputenc}
\usepackage{amsmath}
\usepackage{amsfonts}
\usepackage{float}
\usepackage{graphicx}

\def\spvecA#1;{\if;#1;\else #1\cr \expandafter \spvecA \fi}

\makeatletter
\renewcommand*{\@fnsymbol}[1]{\ifcase#1\or*\or**\else\@arabic{#1}\fi}
\makeatother  
\title{Infinite lattice models by an expansion with a non-Gaussian initial approximation}
\author{
Aleksandr Ivanov\thanks{ivanov.as@physics.msu.ru}\\
M. V. Lomonosov Moscow State University, Faculty of Physics,\\ Leninskie Gory, 119991, Moscow, Russia\\
Institute for Nuclear Research RAS,\\ 60-letiya Oktyabrya prospekt 7a, 117312, Moscow, Russia\\~\\
Vasily Sazonov\thanks{vasily.sazonov@gmail.com}\\
Laboratoire de Physique Th\'eorique, CNRS UMR 8627,\\ 
Universit\'e Paris XI,  F-91405 Orsay Cedex, France}

\begin{document}

\maketitle

\begin{abstract}
Recently, a convergent series employing a non-Gaussian initial approximation
was constructed and shown to be 
an effective computational tool for the finite size lattice models with a polynomial interaction.
Here we show that the Borel summability is a sufficient condition for the correctness of the 
convergent series applied to infinite lattice models. We test the numerical workability of the convergent series method
by examine one- and two-dimensional $\phi^4$-infinite lattice models.
The comparison of the convergent series computations
and the infinite lattice extrapolations of the Monte Carlo simulations reveals an agreement
between two approaches.
\end{abstract}

\section{Introduction}
The development of effective computational methods for systems with a large
number of degrees of freedom (d. o. f.) is the one of the main open problems in the modern theoretical physics.
Conventionally, one uses Monte Carlo simulations to handle the computational complexity,
rising with a growth of the number of particles, lattice sites, etc. However, the Monte Carlo
method has two important limitations. First, the computations with an infinite amount of d. o. f. 
(what for lattice models is equivalent to an infinite volume limit) 
are accessible only as results of an extrapolation procedure. Second, Monte Carlo simulations are based on the
probabilistic interpretations of the Boltzmann weight and, therefore, are not applicable to systems,
described by complex actions. In the current work, we continue studying the convergent series (CS) method 
\cite{Halliday, Ushveridze1983, CSLatticePhi4}, 
which provides new possible ways for bypassing the sign problem \cite{ComplexCS} and for accessing the direct computations in the infinite volume limit, 
focusing on the latter issue.

The divergence of the standard perturbation theory (SPT) is caused by the incorrect interchange of the summation and integration
due to the inaccurate account of large fluctuations of fields. In another words,
the exponent of the polynomial interaction being expanded into the Taylor series, grows to fast at large fields with respect to the Gaussian initial approximation. 
Within the framework of CS the later problem is treated
by choosing a non-standard interacting initial approximation, providing a sufficient decay for competing with a growth
of the expanded part of the Boltzmann weight.
\footnote{For alternative methods see \cite{MeuricePRL, Meurice2004, Belokurov1, SazonovFermions, HowToResum, ConstrMHOI}.}

Initially, the convergent series method was proposed for the quantum anharmonic oscillator and scalar field theories 
\cite{Halliday, Ushveridze1983, Shaverdyan1983, UshveridzeSuper}.
Later, different aspects of the method, including the RG-analysis and strong coupling expansion, 
were developed in \cite{Turbiner, SISSAKIAN1992, VPTSolovtsov, Nalimov}. 
In all these earlier constructions the applicability of the dimensional regularization \cite{Leibbrandt}
to handle the limit of the infinite number of d. o. f. was assumed.
However, as we show in the current paper, in some cases the dimensional regularization may affect main underlying principles of the CS method
and, thus, further mathematical studies are needed.

In recent works \cite{IvanovProc, CSLatticePhi4, ComplexCS} the convergent series 
for the real and certain complex action models defined 
on finite lattices was constructed in a rigorous way. It was shown there that the application of the
dimensional regularization can be interpreted as an additional re-summation 
procedure accelerating the series convergence. 

In this paper we describe in details the problems related to the utilization of the dimensional
regularization and prove that the Borel summability  is a sufficient (but not a necessary one) condition for justifying 
the re-summation procedure substituting the dimensional regularization. 
To stress a re-summation character of handling the limit of an infinite amount of d. o.f., we numerically investigate 
the applicability of the CS method to infinite systems where the dimensional regularization cannot be utilized at all. 
Namely, we consider the $\phi^4$-lattice model on the infinite lattice in one and two dimensions.


%
%

\section{CS method and dimensional regularization}
\label{CSM}
We start by describing the main steps of the convergent series construction and by specifying
problems related to the dimensional regularization.
For this purpose we consider a partition function of the one-dimensional $\phi^4$-model,
\begin{equation}
  Z := \frac{1}{Z_0} \int {\cal D}\phi\, e^{-S[\phi]}\,,
\label{Zao}
\end{equation}
with $Z_0$ being a partition function of the free theory and the action given by
\begin{equation}\label{phi4}
  S[\phi] = \int d x \Bigl ( \frac{1}{2} (\partial _ x \phi) ^ 2 - \frac{1}{2} m ^ 2 \phi ^ 2 - \frac{\lambda}{4!} \phi ^ 4 \Bigr ),
\end{equation}
where $m$ is the mass and $\lambda$ is the coupling constant.

The main idea of the CS method is to generate an absolutely convergent series by expanding the integrand
into the series with positive terms only. Then, the interchange of the summation and integration cannot affect the absolute 
convergence and if the initial quantity was finite, the series must converge. To implement this strategy within the CS method,
one chooses a non-standard initial approximation given by a non-local action of the form
\begin{equation}
N[\phi] = S_2[\phi] + \sigma S_2[\phi]^2\,,
\label{Nact}
\end{equation}
where
\begin{equation}
  S_2[\phi] = \int d x \Bigl ( \frac{1}{2} (\partial_x \phi) ^ 2 - \frac{1}{2} m ^ 2 \phi ^ 2\Bigr)
\label{S2}
\end{equation}
is the quadratic part of the action \eqref{phi4}.
Then, the Sobolev's inequalities guarantee the existence of the parameter $\sigma$ and provide its minimal value \cite{Talenti} such that
\begin{equation}
  \sigma S_2[\phi]^2 \geq S_{I}[\phi]~~ \Longleftrightarrow~~ N[\phi] \geq S[\phi]\,,
\label{mainineq}
\end{equation}
where the interaction part of the action is defined as $S_I[\phi] = S[\phi] - S_2[\phi]$.

Splitting the action into the non-perturbed and perturbation parts as $S[\phi] = N[\phi] + (S[\phi] - N[\phi])$ 
we obtain
\begin{equation}
  Z = \frac{1}{Z_0} \int {\cal D}\phi\, e^{-N[\phi]} \sum_k \frac{(N[\phi] - S[\phi])^k}{k!}\,.
\end{equation}
When the inequality \eqref{mainineq} is satisfied, the expression $e^{-N[\phi]} (N[\phi] - S[\phi])^k \geq 0$
and one can interchange the summation and integration
\begin{equation}
  Z = \frac{1}{Z_0} \sum_k \frac{1}{k!} \int {\cal D}\phi\, e^{-N[\phi]} (N[\phi] - S[\phi])^k\,.
\label{exp1}
\end{equation}
Indeed, 
\begin{eqnarray}
\nonumber
  |Z| \leq \frac{1}{Z_0} \sum_k \frac{1}{k!} \int {\cal D}\phi\, e^{-N[\phi]} \Big|(N[\phi] - S[\phi])^k\Big| = \\
  \frac{1}{Z_0} \int {\cal D}\phi\, e^{-N[\phi]} \sum_k \frac{1}{k!} (N[\phi] - S[\phi])^k = Z\,,
\end{eqnarray}
so the convergence of the series is guaranteed by the assumption $Z < \infty$.

The computation of terms in \eqref{exp1} can be done by several equivalent ways \cite{Shaverdyan1983, UshveridzeSuper, SISSAKIAN1992}, facing similar difficulties requiring the utilization
of the dimensional regularization, here we follow \cite{UshveridzeSuper}. The equation \eqref{exp1}
can be exactly rewritten as
\begin{eqnarray}
\nonumber
  Z &=& \frac{1}{Z_0} \sum_k \frac{1}{k!} \int {\cal D}\phi\, e^{-S_2[\phi] - \sigma S_2^2[\phi]} (\sigma S_2^2[\phi] - S_I[\phi])^k \\
  &=& \frac{1}{Z_0} \sum_k \frac{1}{k!} \int {\cal D}\phi\, \int_{0}^\infty dt\, \delta(t - \sqrt{S_2[\phi]}) 
  e^{-t^2 - \sigma t^4} (\sigma t^4 - S_I[\phi])^k \,.
\label{exp2}
\end{eqnarray}
The next step is the interchange of integrations over the axillary variable $t$ and field $\phi$.
It can be easily justified only for systems with a finite amount of d. o. f., 
i.e. when instead of the path integral one deals with a finite dimensional integral.
Yet, we proceed with a continuum case with infinitely many d.o.f. Formally denoting $V := \int dx$, we perform
the rescaling of variables, $\phi \rightarrow t\phi$.
\begin{eqnarray}
\nonumber
  Z  &\simeq& \frac{1}{Z_0} \sum_k \frac{1}{k!} \int_{0}^\infty dt\, \int {\cal D}\phi\, t^V\, \delta(t - t \sqrt{S_2[\phi]}) 
  e^{-t^2 - \sigma t^4} (\sigma t^4 - t^4 S_I[\phi])^k \\
\nonumber
  &=& \frac{1}{Z_0} \sum_k \frac{1}{k!} \int_{0}^\infty dt\, t^{V + 4k - 1}\,e^{-t^2 - \sigma t^4} \int {\cal D}\phi\, \delta(1 - \sqrt{S_2[\phi]}) 
  (\sigma - S_I[\phi])^k\\
\nonumber
  &=& \frac{1}{Z_0} \sum_k \frac{1}{k!} \int_{0}^\infty dt\, t^{V + 4k - 1}\,e^{-t^2 - \sigma t^4} \cdot \\
  &\cdot&\int {\cal D}\phi\, \delta(1 - \sqrt{S_2[\phi]}) \sum_{l=0}^k \binom{k}{l} \sigma^{k - l} (-S_I[\phi])^l\,,
\label{exp3}
\end{eqnarray}
where the sign '$\simeq$' indicates a non-strict equality. 
The dimensional regularization prescribes $V = 0$. From the first look it seems reasonable,
at least, the dimensional regularization is a common tool in the quantum field theory and at this step it does not violate the main principle of the CS method: the positivity of
the series terms. However, the equation \eqref{exp3} is not a final answer and the main difficulties show up later on. In the following
we do not apply dimensional regularization, but instead we keep '$V$' in formulas explicitly for clarifying the appearing problems.

All the steps carried out for the $\phi^4$-model and leading to \eqref{exp3} can be repeated for any correlation function of 
the free theory with the Gaussian action $S_2[\phi]$
\begin{eqnarray}
\nonumber
  &&\int {\cal D}\phi\,\phi(x_1)...\phi(x_Q)\, e^{-S_2[\phi]}\\
\nonumber
  &&=\, \int_{0}^\infty dt\, t^{V + Q - 1}\,e^{-t^2} \int {\cal D}\phi\, \phi(x_1)...\phi(x_Q)\, \delta(1 - \sqrt{S_2[\phi]})\\
  &&=\, \frac{1}{2} \Gamma\Big(\frac{V + Q}{2}\Big) \int {\cal D}\phi\, \phi(x_1)...\phi(x_Q)\, \delta(1 - \sqrt{S_2[\phi]})\,.
\label{ident}
\end{eqnarray}
Then, using \eqref{ident}, each term in the double sum \eqref{exp3} can be rewritten as a moment of a Gaussian distribution, i.e.
the convergent series is expressed as a re-summation of the standard perturbation theory
\begin{eqnarray}
  Z  &=&  \frac{1}{Z_0} \sum_k \frac{1}{k!} \int_{0}^\infty dt\, t^{V + 4k - 1}\,e^{-t^2 - \sigma t^4} 
  \sum_{l=0}^k \binom{k}{l} \frac{\sigma^{k - l}}{\Gamma\Big(\frac{V + 4l}{2}\Big)}  f_l\,,
\label{exp4}
\end{eqnarray}
where
\begin{eqnarray}
  f_l &:=& \int {\cal D}\phi\, e^{-S_2[\phi]}  (-S_I[\phi])^l\,.
\end{eqnarray}
The only problem of \eqref{exp4} is the explicit dependence on $V$. Even for the one-dimensional system under the consideration, where the UV-renormalization is not needed,
the formulation \eqref{exp4} requires regularization. In \cite{CSLatticePhi4} we have proved the existence 
of an internal perturbative symmetry of the CS-method allowing the change of $V$ to $V + \tau$ in the formula \eqref{exp4} for an arbitrary $\tau$.
Namely, the perturbative expansions of \eqref{exp4} with $V$ and of \eqref{exp4} with $V + \tau$ coincide.
Consequently, the utilization of the dimensional regularization is correct at least perturbatively up to all loops.
At the same time, any modification of the true infinite value of $V$ changes relevant weights of different terms in the sum over '$l$' in \eqref{exp4},
what may lead to the loss of the positivity of the series terms.

\begin{figure*}
\centering 
\includegraphics[height=5.5cm]{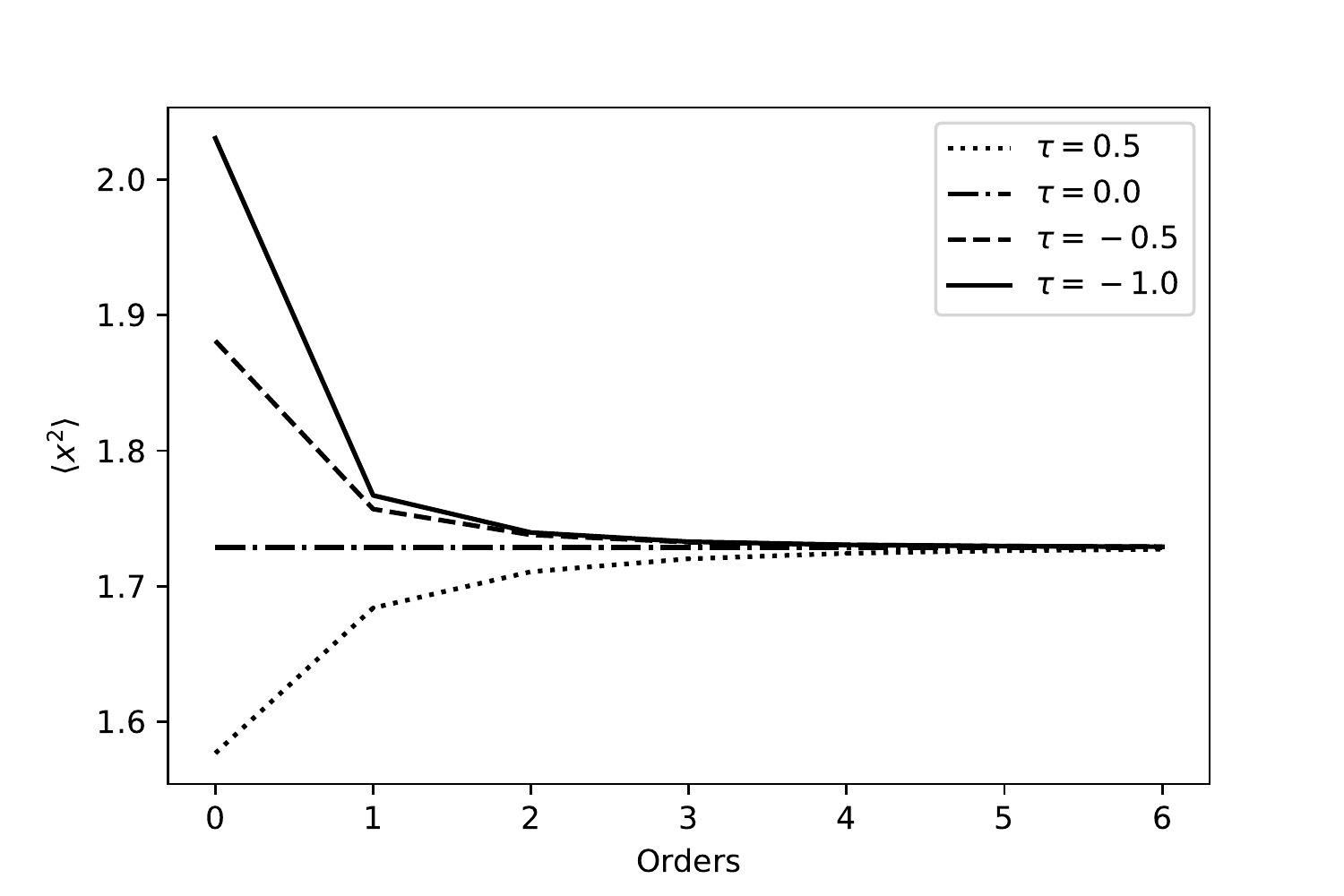}
\caption{Dependence of $I$ on the number of CS terms at different $\tau$.}
\label{tauDep}
\end{figure*}

It is hard (or may be impossible) to demonstrate such an effect on the quantum field theory example,
however one can show the possibility of such a situation referring to the finite dimensional case. 
Consider an one-dimensional integral
\begin{equation}
  I := \int_{-\infty}^{\infty} dx\,x^2\,e^{-\frac{1}{2} m^2 x^2 - \frac{\lambda}{4!} x^4}\,,
\end{equation}
the fastest convergence of CS corresponding to this integral can be achieved by choosing 
$\sigma = \frac{\lambda}{6 m^4}$. Then, the whole value of the integral $I$ is given by the zero CS approximation
(it coincides with the integral) and all other corrections are zero. However, the integral $I$ can be still
formally written as \eqref{exp4}
\begin{eqnarray}
\nonumber
  I  &=& \sum_k \frac{1}{k!} \int_{0}^\infty dt\, t^{V + \tau + 4k + 1}\,e^{-t^2 - \sigma t^4} \cdot \\
  &&\cdot \sum_{l=0}^k \binom{k}{l} \frac{1}{\Gamma\Big(\frac{V + \tau + 4l}{2}\Big)} \int dx\,x^2\, e^{-\frac{1}{2} m^2 x^2} \sigma^{k - l} \big(-\frac{\lambda}{4!} x^4\big)^l\,.
\label{I1}
\end{eqnarray}
When $\tau = 0$, the zero term coincides with $I$ and all corrections are zero, the latter happens due to the fact that
all integrands in this case are effectively proportional to $(\sigma(\frac{1}{2} m^2 x^2)^2-\frac{\lambda}{4!} x^4)^k = 0$ when $k \neq 0$.
When $\tau \neq 0$, the ratio between terms in the sum over $l$ in \eqref{I1}
is destabilized and they can become negative. This is indeed the case for the
integral \eqref{I1}. We present the dependence of $I$ on $\tau$ and on the number of CS coefficients taken into account in Fig. \ref{tauDep}.
The decrease of the curves at $\tau < 0$ indicates the presence of negative terms in CS.

Despite that in the considered example the CS terms lose their positivity due to the change of $\tau = 0$ to $\tau = -1$ or to $\tau = -1/2$,
one can note that the series nevertheless converges to the correct answer. In the work \cite{CSLatticePhi4} it is explained by
employing an additional regularization of the form
\begin{eqnarray}
\label{limrep}
  I &=& \lim_{\gamma \rightarrow 0} I(\gamma)\,,\\
\label{gammareg}
  I(\gamma) &:=& \int_{-\infty}^\infty dx\, e^{-S[x]}e^{-\gamma S_2^3[x]}\,,
\end{eqnarray}
where the notations $S[x] := \frac{1}{2} m^2 x^2 + \frac{\lambda}{4!} x^4$ and $S_2[x] := \frac{1}{2} m^2 x^2$
are used to emphasize the analogy with lattice models.
The representation \eqref{limrep} is always correct for the finite dimensional integrals, i.e.
there is always such $\gamma_*$, that $I(\gamma_*)$ approximates $I(\gamma)$ with an arbitrary precision.
At $\gamma > 0$ the application of the CS method for computing $I(\gamma > 0)$ generates
the convergent series even at $\tau \neq 0$. The latter can be proved by considering the large order asymptotic
of the perturbative re-expansion of the CS representation for $I(\gamma > 0)$. 

In case of the infinite dimensional integrals, the $\gamma$-regularization \eqref{limrep}, \eqref{gammareg} is not valid. 
It happens because the term analogous to $\gamma S_2^3[x]$ is non-local and the models with local and non-local actions have different scalings at $V \rightarrow \infty$. Nevertheless, note
that the failure of the regularization \eqref{limrep}, \eqref{gammareg} prohibits the proof from \cite{CSLatticePhi4} for infinite volumes, 
but does not necessary imply the inapplicability of the CS method.
Here we study numerically the workability of the CS method investigating 
the $\phi^4$-model defined on the one- and two-dimensional infinite lattices.
Then, we prove the correctness of the results by showing that the Borel summability is a sufficient 
condition for the CS applicability.


\section{Convergent series for the lattice $\phi^4$-model} 
The discretized version of the action \eqref{phi4} (now in an arbitrary dimension $D$) is given by 
\begin{eqnarray}\label{S_E}
  &&S_E = S_2[\varphi] + S_4[\varphi]\,,
\\  
\label{S2d}
  &&S_2[\varphi] = -\frac{1}{2} \sum _ {n, \mu} \varphi_n \varphi _ {n + \mu} + \frac{1}{2}(2 D + M ^ 2) \sum_n \varphi _ n ^ 2\,,
\\
\label{S4}
  &&S_4[\varphi] = \frac{\lambda}{4!} \sum_n \varphi_n^4,
\end{eqnarray}
where $\varphi = a \phi$, $M = a m$, $x = n a$, 
$\int d ^ D x \to a ^ D \sum _ n$, $(\partial _ \mu \phi) ^ 2 \to \frac{1}{a ^ 2} \sum _ \mu (2 \phi (n a) - \phi (n a + \mu a) - \phi (n a - \mu a))$, 
$n$ is the $D$-dimensional integer vector, $\mu$ is the unit vector directed along the $\mu$-axis, $a$ is the lattice spacing.
This action leads to the standard perturbation theory with tetravalent vertices and a bare propagator given by 
\begin{equation}
  \langle \varphi _ n \varphi _ m \rangle_0 = \int _ {-\pi} ^ {\pi} \frac{d ^ D k}{(2 \pi) ^ D} \frac{e ^ {i k (n - m)}}{4 \sum _ \mu \sin ^ 2 \frac{k _ \mu}{2} + M ^ 2}.
\end{equation}
The CS expression for the two-point function of the $\phi^4$-model on the lattice of the size $V$ is derived in a complete analogy to \eqref{exp4} 
\begin{eqnarray}
  \langle \varphi_n \varphi_m \rangle_\lambda &=& \sum_{k = 0}^\infty \frac{1}{k!} 
  \int_0^\infty\,dt\, t^{V + \tau + 4 k + 1} \exp \Bigl (-t ^ 2 - \sigma t ^ 4 \Bigr)  \nonumber\\
  &&\cdot\sum_{l=0}^k \binom{k}{l} \sigma ^ {k - l} \frac{2 f_l(n, m) \lambda ^ l}{\Gamma \Bigl (\frac{V + \tau + 4l + 2}{2}\Bigr)}, 
\label{compform}
\end{eqnarray}
where $f_l(n, m) / (l!)$ are the coefficients of the standard perturbation theory, i.e.
$\langle \varphi _ n \varphi _ m \rangle _ \lambda \sim  \sum_{l = 0}^\infty f_l(n, m) / (l!) \lambda^l$ 
and $\sigma$ is the parameter of the new initial approximation, we choose it to be the minimal $\sigma$ satisfying the inequality \eqref{mainineq},
\begin{equation}
  \sigma = \frac{\lambda}{6 M^4}\,.
\end{equation}
We apply \eqref{compform} to the limiting case $V + \tau = 0$, $V \rightarrow \infty$ and compare
the results with corresponding extrapolations of the path integral Monte Carlo (PIMC) simulations  
data and the Borel re-summation of the standard perturbation theory.

\section{Numerical results}
The main building blocks of the CS are the terms of the standard perturbation theory.
Their computations in the infinite volume limit are much simpler than for finite lattices. 
At finite lattice volumes, an evaluation of each diagram of the $\phi^4$-model involves the multiplication of $4$ bare-propagator
matrices of the size $V \times V$ per vertex leading to the enormously long computations.
However, by turning to the computations in the infinite volume limit,
one substitutes tedious matrix multiplications by numerical integrations, which can be performed with the Monte Carlo technique.

\begin{figure*}
\centering 
\includegraphics[height=5.5cm]{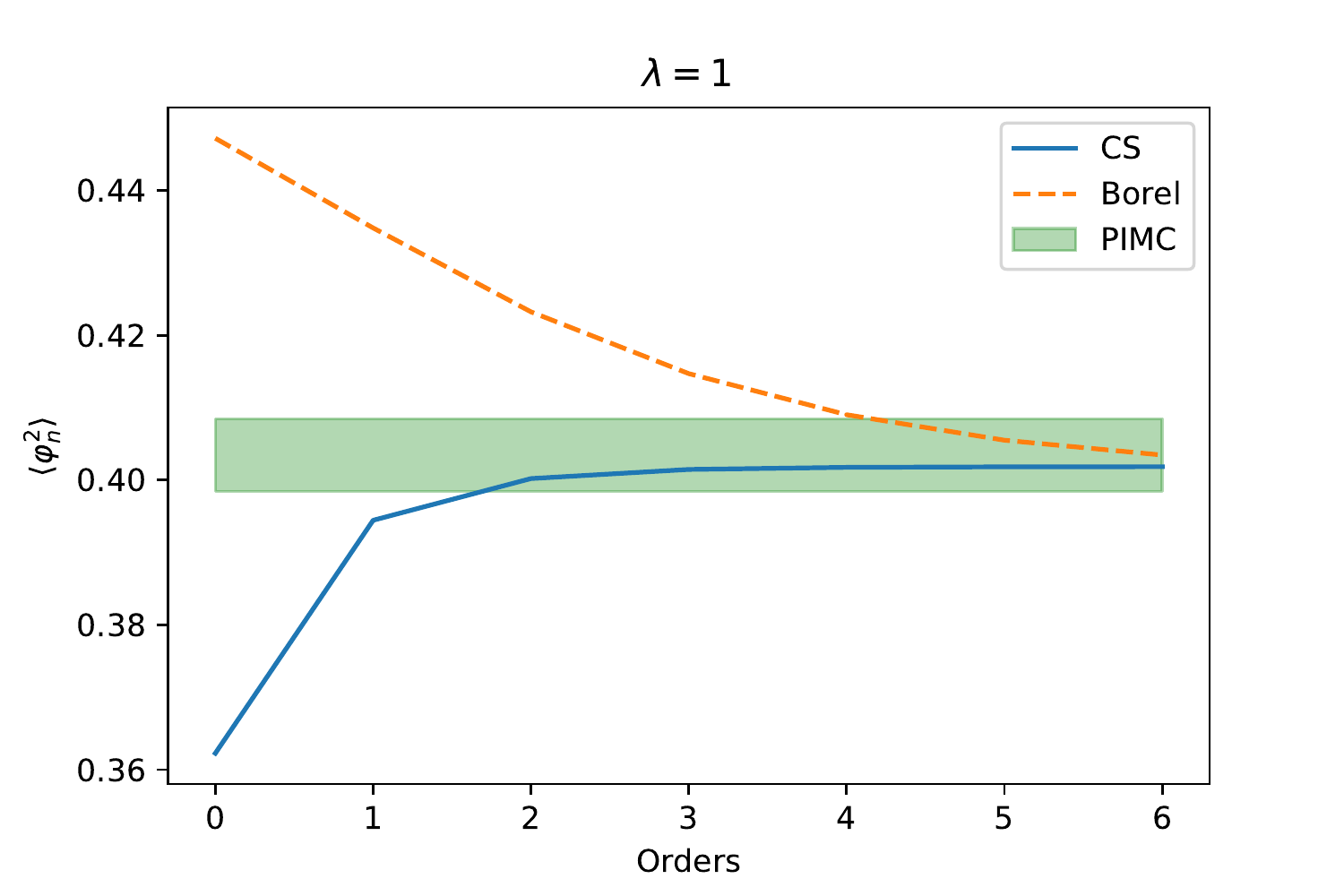}\par\medskip
\includegraphics[height=5.5cm]{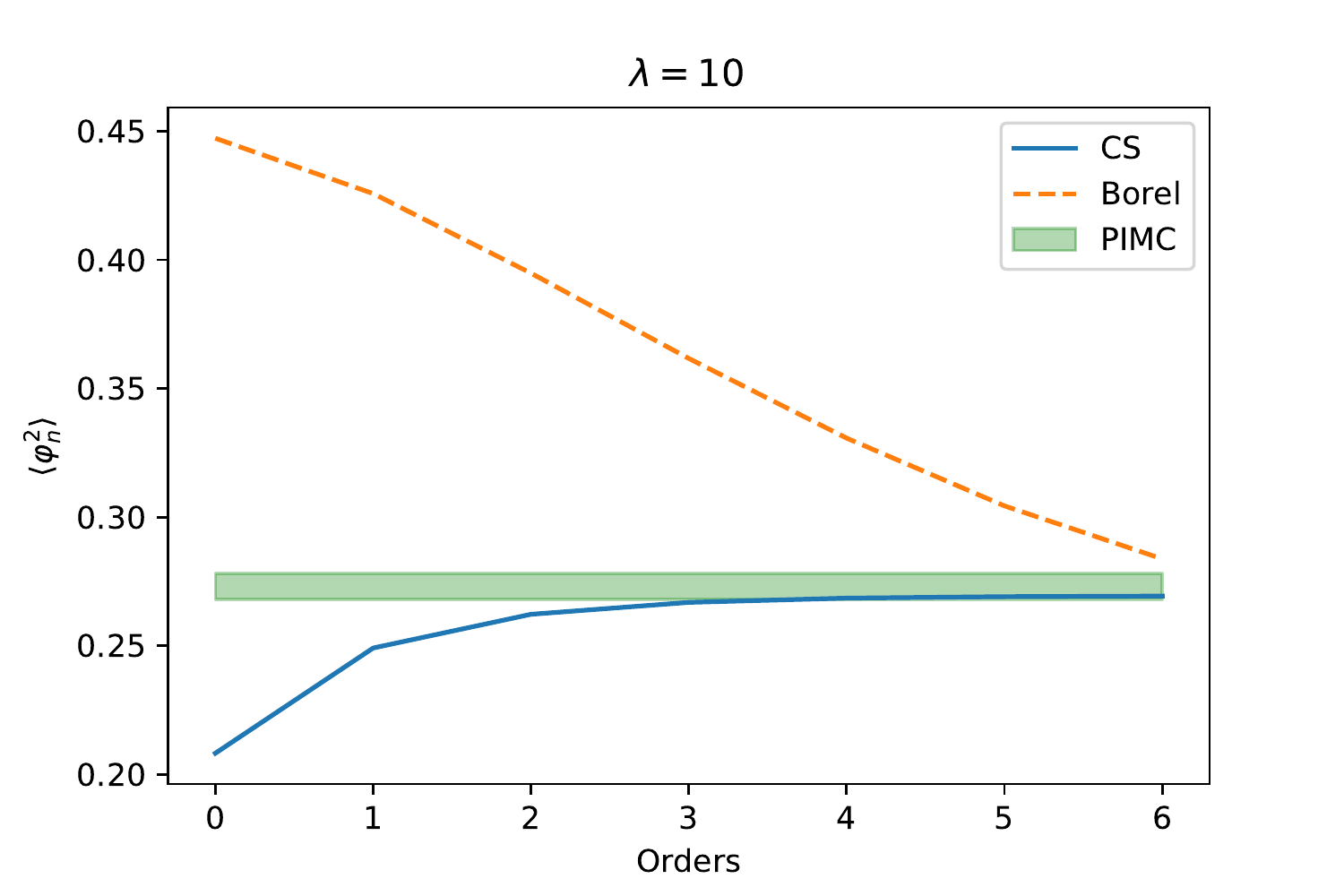}
\caption{Dependence of $\langle\varphi_n^2\rangle$ computed with the CS method and Borel re-summation on the number of series terms 
and comparison with the extrapolated PIMC data, $D = 1$, at $\lambda = 1$ and $\lambda = 10$.}
\label{Comparison}
\end{figure*}

\begin{figure*}
\centering 
\includegraphics[height=5.5cm]{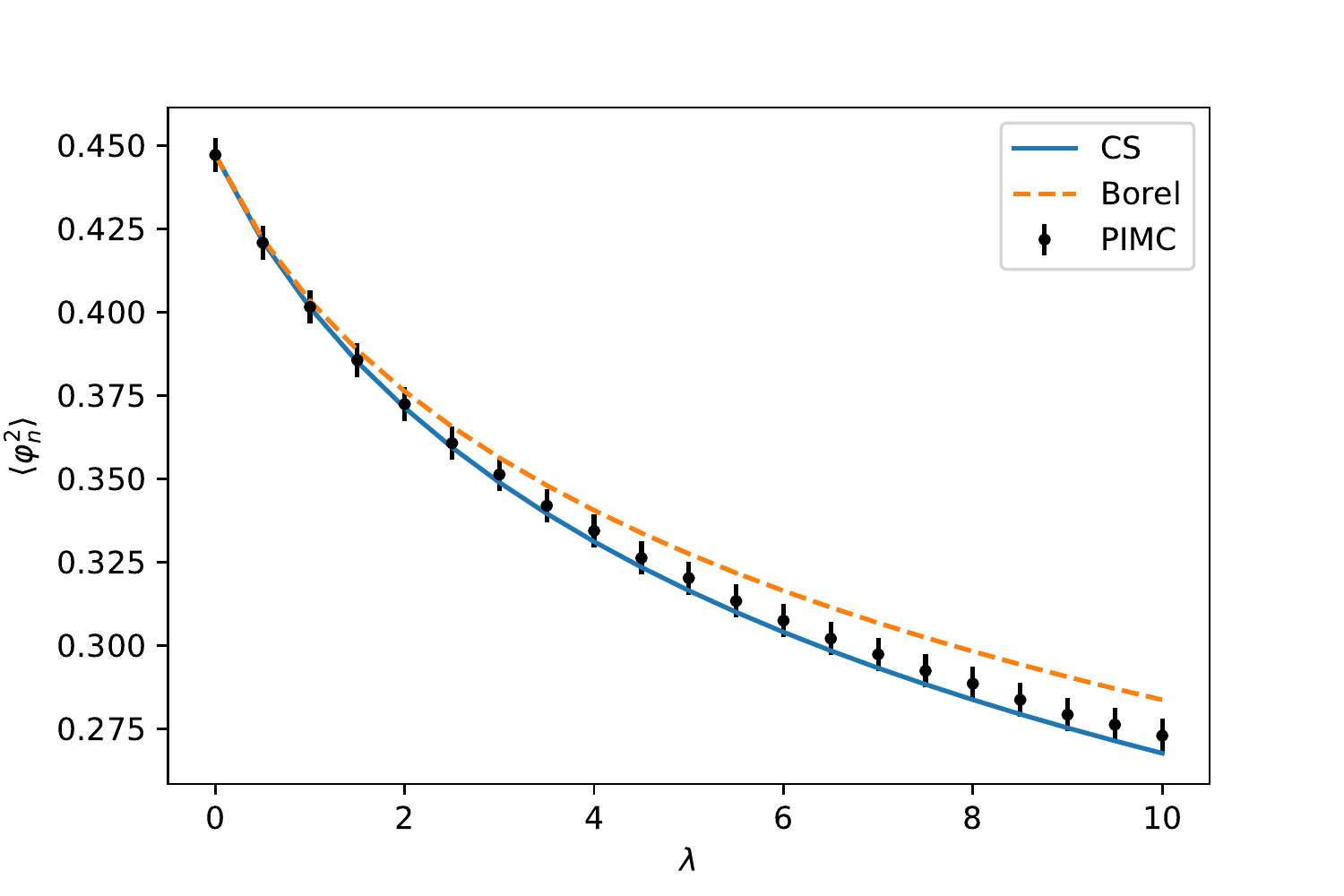}
\caption{Dependence on the coupling constant of the $\langle\varphi_n^2\rangle$-operator computed with the sixth order of the CS and Borel series and
PIMC extrapolation, $D = 1$.}
\label{3methods}
\end{figure*}

On the Fig. \ref{Comparison} we present the $\langle\varphi_n^2\rangle$-operator of the one-dimensional lattice $\phi^4$-model depending 
on the number of computed terms of the standard perturbation theory at
two reference values of the coupling constant, $\lambda = 1$ and $\lambda = 10$.
All plots exhibit that the CS method converges to the extrapolated Monte Carlo answer faster 
than the Borel re-summation procedure (for the details of the PIMC computations and Borel re-summation see Appendix \ref{APIMCBorel}).
In Fig. \ref{3methods} we show the 
coupling constant dependence of the $\langle\varphi_n^2\rangle$-operator computed with six orders of
the CS and Borel series. The reference 'correct' answer is obtained by the extrapolation of the PIMC computations.
The CS method demonstrates better precision at large values of the coupling constant than Borel re-summation.

\begin{figure*}
\centering 
\includegraphics[height=5.5cm]{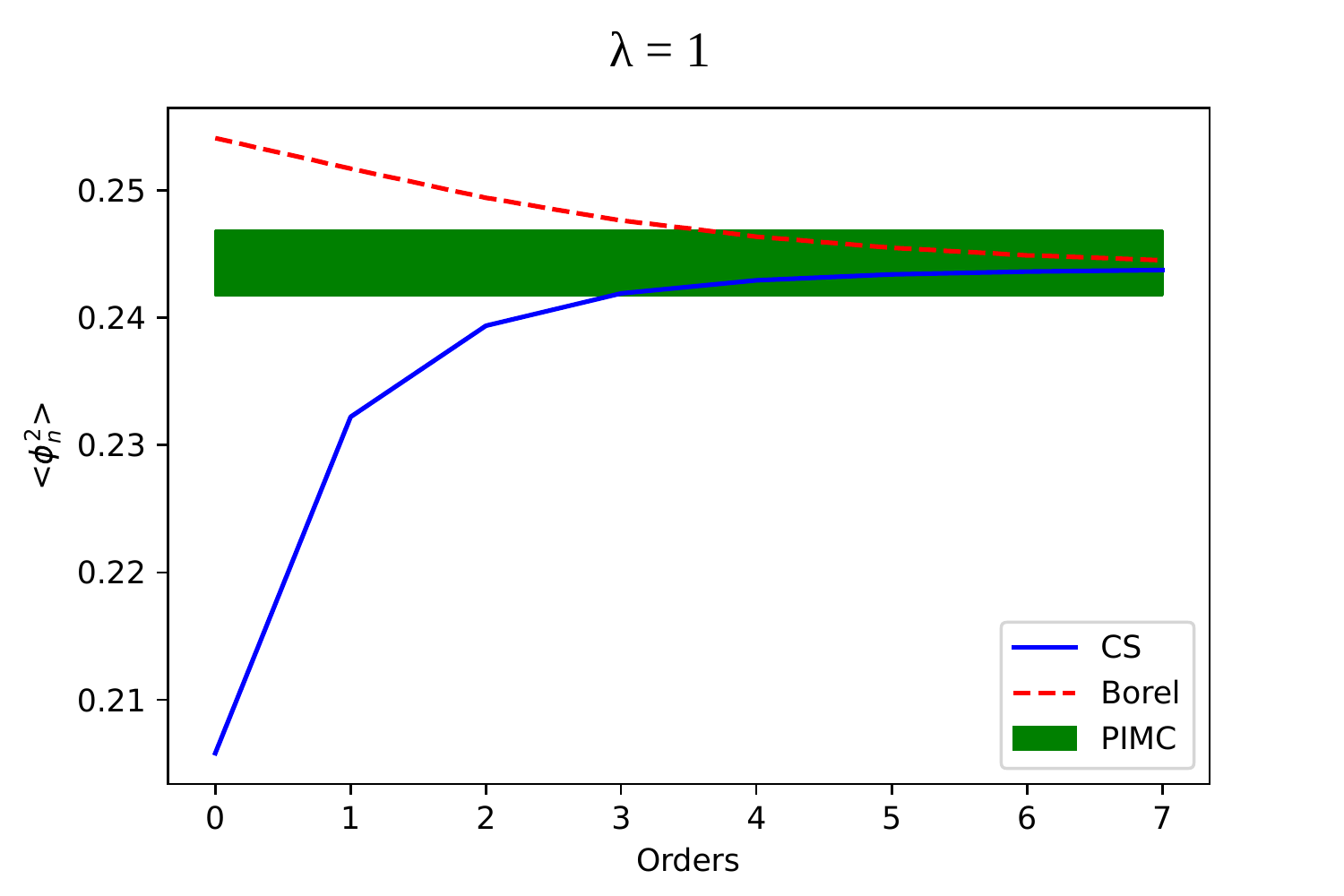}\par\medskip
\includegraphics[height=5.5cm]{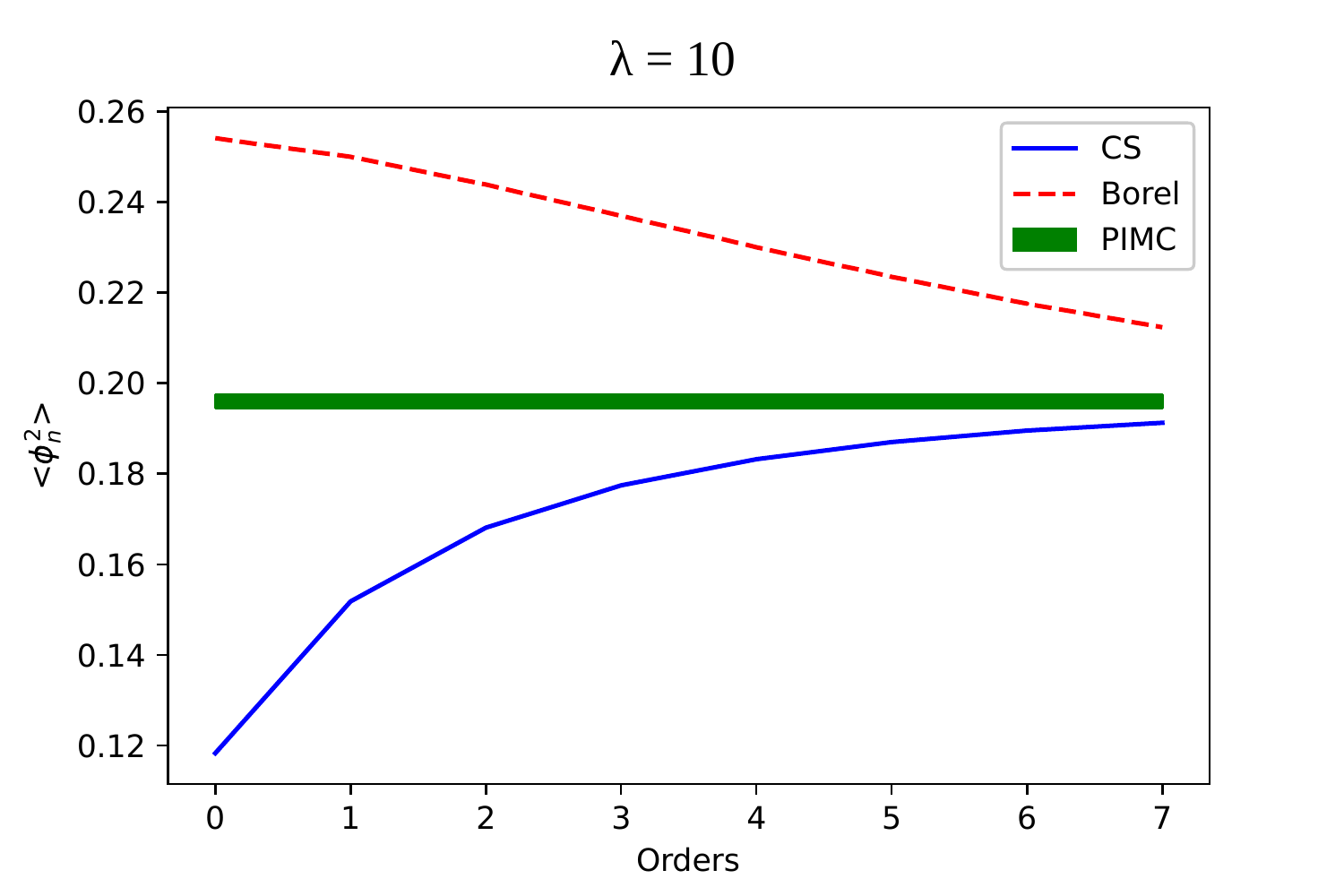}
\caption{Dependence of $\langle\varphi_n^2\rangle$ computed with the CS method and Borel re-summation on the number of series terms 
and comparison with the extrapolated PIMC data, $D = 2$, at $\lambda = 1$ and $\lambda = 10$.}
\label{Comparison2}
\end{figure*}

The consideration of the two-dimensional infinite lattice demonstrates similar results.
On the Fig. \ref{Comparison2} we present the results for the $\langle\varphi_n^2\rangle$-operator depending 
on the number of computed terms of the standard perturbation theory at $\lambda = 1$ and $\lambda = 10$.
\begin{figure*}
\centering 
\includegraphics[height=5.5cm]{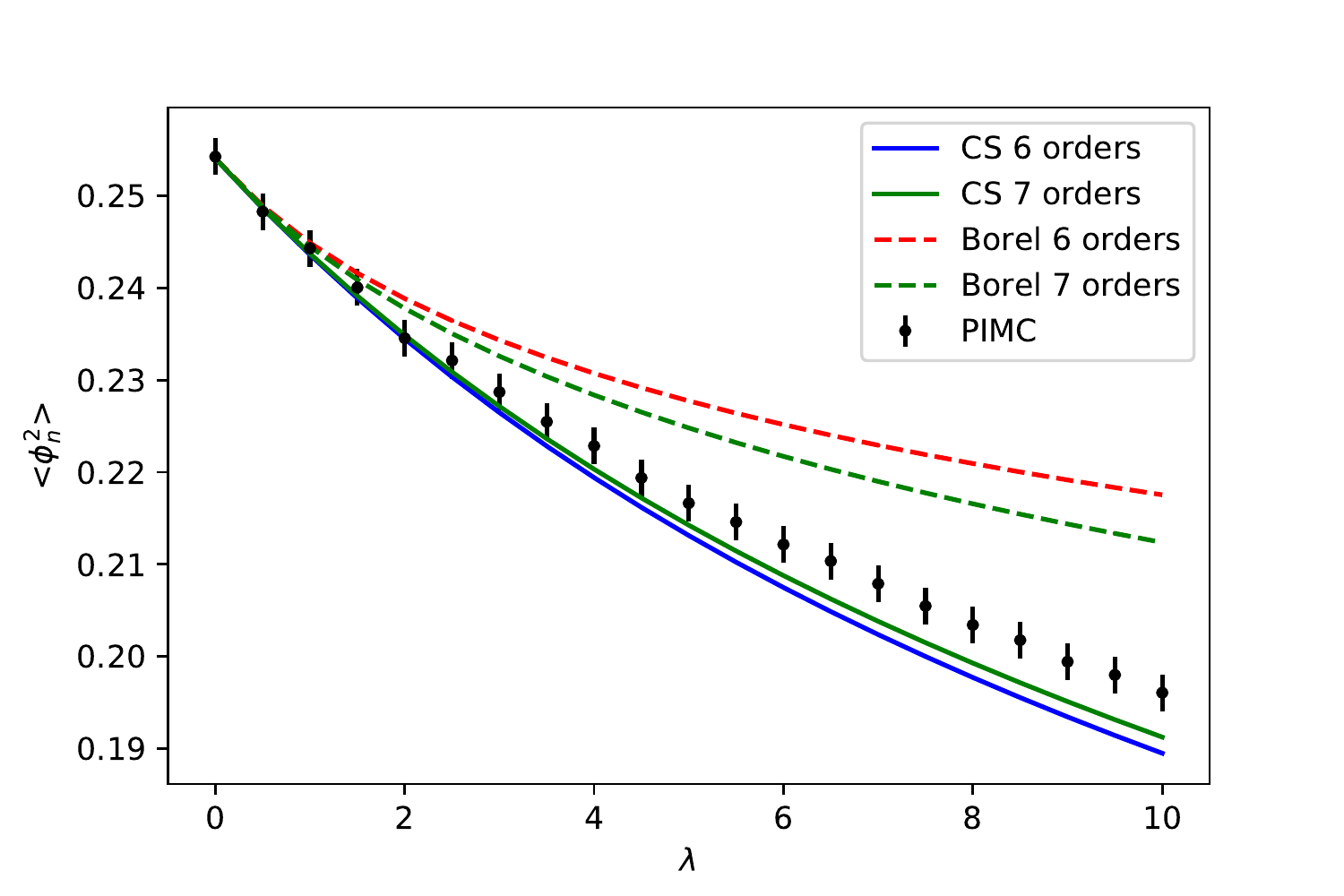}
\caption{$\langle\varphi_n^2\rangle$-operator depending on the coupling constant computed with the sixth and seventh orders of the CS and Borel series and
PIMC extrapolation, $D = 2$.}
\label{3methods2}
\end{figure*}
In Fig. \ref{3methods2} we show the value of the $\langle\varphi_n^2\rangle$-operator depending on the coupling constant, 
emphasizing the improvement achieved by going from the sixth order approximation to the seventh.

\section{Relation to the Borel summability}
The remarkable workability of the CS method for finite lattices was explained in \cite{CSLatticePhi4}
by considering a $\phi^4$-model regularized by an additional non-local interaction with a vanishing coupling. As it was
mentioned in the Section \ref{CSM}, such approach can not be used for the infinite lattices, since
the regularization used in \cite{CSLatticePhi4} adds a non-local term to the action.
Models with non-local interactions have an unusual (different from ones with a local interaction) scaling with respect to the lattice volume.
Consequently, non-local models never approximate local ones in the limit of the infinite amount of d. o. f.
Nevertheless, we use the regularization of the $\phi^4$-model of the type of \cite{CSLatticePhi4} just as a motivation
for an another construction, providing a link between the applicability of the CS method for infinite lattices and
the Borel summability.

Let us consider a model, defined on the infinite lattice by the action
\begin{equation}
  S_\gamma = S_2[\varphi] + S_4[\varphi] + \gamma (S_2[\varphi])^3,
\label{Sgamma}
\end{equation}
where $S_2[\varphi]$ and $S_4[\varphi]$ are given by \eqref{S2} and \eqref{S4} respectively and $\gamma \geq 0$ is a parameter. 
For this model one can write a formal expression analogous to \eqref{exp4}
\begin{eqnarray}
  Z_\gamma  =  \frac{1}{Z_0} \sum_k \frac{1}{k!} \int_{0}^\infty dt\, t^{V + 4k - 1}\,e^{-t^2 - \sigma t^4 - \gamma t^6}
  \sum_{l=0}^k \binom{k}{l} \frac{\sigma^{k - l} f_l}{\Gamma\Big(\frac{V + 4l}{2}\Big)}\,.
\label{Zgamma1}
\end{eqnarray}
The equation \eqref{Zgamma1} is ill defined when $V \rightarrow \infty$. However, as it was shon in \cite{CSLatticePhi4}, at $\gamma = 0$ 
for any $V$, there is an invariance (at least perturbative) allowing one to substitute $V$ by $V + \tau = w$, where $w$ is finite. 
In case of finite lattices \cite{CSLatticePhi4},
the continuity of $Z_\gamma$ at $\gamma \rightarrow 0$ was used for justifying the change from $V$ to $V + \tau = w$ in the non-perturbative sense. 
Since limits $V\rightarrow\infty$ and $\gamma\rightarrow 0$ do not commute, we cannot follow the same way.
Instead of thinking about the regularization of the model, we just consider the series \eqref{Zgamma1}. 
We do not perform the limit $V \rightarrow \infty$. Instead, we do a formal substitution $V --> w$
(as it is prescribed by the CS method) and prove that it leads to the correct answer, independently on $V$, if $Z$ is Borel-summable.
Indeed, considering the asymptotic of large orders \cite{CSLatticePhi4}, one can show that for $\gamma > 0$ the series
\begin{eqnarray}
  Z_\gamma^w  :=  \frac{1}{Z_0} \sum_k \frac{1}{k!} \int_{0}^\infty dt\, t^{w + 4k - 1}\,e^{-t^2 - \sigma t^4 - \gamma t^6}
  \sum_{l=0}^k \binom{k}{l} \frac{\sigma^{k - l} f_l}{\Gamma\Big(\frac{w + 4l}{2}\Big)}\,.
\label{Zgamma2}
\end{eqnarray}
is absolutely convergent. The re-expansion (re-summation) of \eqref{Zgamma2} 
in the coupling constant $g$ generates an another absolutely convergent series for $Z_\gamma^w$,
\begin{eqnarray}
  Z_\gamma^w  &=&  \frac{1}{Z_0} \sum_{l=0}^\infty \frac{1}{l!} \int_{0}^\infty dt\, t^{w + 4l - 1}\,e^{-t^2 - \gamma t^6}
  \frac{f_l}{\Gamma\Big(\frac{w + 4l}{2}\Big)}\,.
\label{Zgamma3}
\end{eqnarray}
The equivalence of \eqref{Zgamma2} and \eqref{Zgamma3} is guarantied by their absolute convergence.
Due to the absolute convergence, one can interchange the summation and integration in \eqref{Zgamma3}
\begin{eqnarray}
  Z_\gamma^w  &=&  \frac{1}{Z_0} \int_{0}^\infty dt\, t^{w - 1}\,e^{-t^2 - \gamma t^6}
  \sum_{l=0}^\infty \frac{t^{4l} f_l/(l!)}{\Gamma\Big(\frac{w + 4l}{2}\Big)}\,,
\label{Zgamma4}
\end{eqnarray}
where $f_l/(l!)$ are the coefficients of the standard perturbation theory.
At $\gamma = 0$ the expression \eqref{Zgamma4} is nothing more than the Borel-Leroy summation of the SPT series for $Z$.
Then, the possibility to take the limit $\gamma \rightarrow 0$ (or in other words to use CS method for the infinite lattices) 
follows from the Borel summability. 

\section{Concluding remarks}
In this paper we have studied the applicability of the convergent series to the systems with an infinite amount of degrees of freedom.
Considering $\phi^4$-model on the one and two-dimensional infinite lattices,
we have provided a numerical evidence of the robustness of the CS method.
Our computations indicate, that the convergent series approaches to the reference Monte Carlo results
(where an infinite lattice volume is reached by an extrapolation procedure) faster than the corresponding
Borel summation.

For a long time (starting from the first papers in the 1980s \cite{Ushveridze1983, Shaverdyan1983, UshveridzeSuper})
the derivation of the method was based on the utilization of the dimensional regularization and
the justification of the CS method for the infinite systems was remaining an open question.
Here we have shown that the utilization of the dimensional regularization is equivalent to an additional re-summation procedure.
We have proved that this additional re-summation preserves the convergence of the CS method if the model
is Borel summable. The Borel summability is a sufficient condition for the correctness of the CS method
in the limit of an infinite amount of degrees of freedom. It is not hard to find an effective action, 
satisfying the inequality analogous to \eqref{mainineq} even in case of some Borel non-summable models,
for instance, for the double-well potential. Then, the logic of the paper \cite{CSLatticePhi4}
suggests, that on the finite lattices the CS method should be correct also for the models which are Borel non-summable.
Taking into account results of the current paper, one can hope that the CS method is applicable to a wider
class of the quantum field theories than the Borel re-summation and can serve as a new tool for the analytic non-perturbative
computations.

\subsection*{Acknowledgments}
The work of AI was supported by the Foundation for the Advancement of Theoretical Physics “BASIS” 17-21-103-1 
and by Russian Science Foundation Grant No. 14-22-00161.
VS acknowledges the support provided by the Austrian Science Fund (FWF) trough the Erwin Schr\"odinger fellowship J-3981.

\appendix
\section{}
\label{APIMCBorel}
\subsection{Borel re-summation}
To perform the Borel re-summation procedure, we use the conformal mapping \cite{Zinn1977} for the analytic continuation 
in the Borel plane. The conformal mapping can be done if the parameter $a$ from
the asymptotic of the high orders of the perturbation theory,
\begin{equation}
\frac{f_k}{k!} \sim (-1)^k \sqrt{2\pi} e \big(\frac{a}{e}\big)^k k^{k + b_0 + 1/2}\,,
\end{equation}
is known. We estimate $a$ using
the values presented in \cite{Zinn2010} for the continuous one- and two-dimensional $\phi^4$-model
\begin{eqnarray}
d = 1\,,~~~ a = 1/8\,,\\
d = 2\,,~~~ a = 1/35.102\, . . . \,.
\label{aparam}
\end{eqnarray}

\subsection{PIMC}
\begin{figure*}
\centering 
\includegraphics[height=5.5cm]{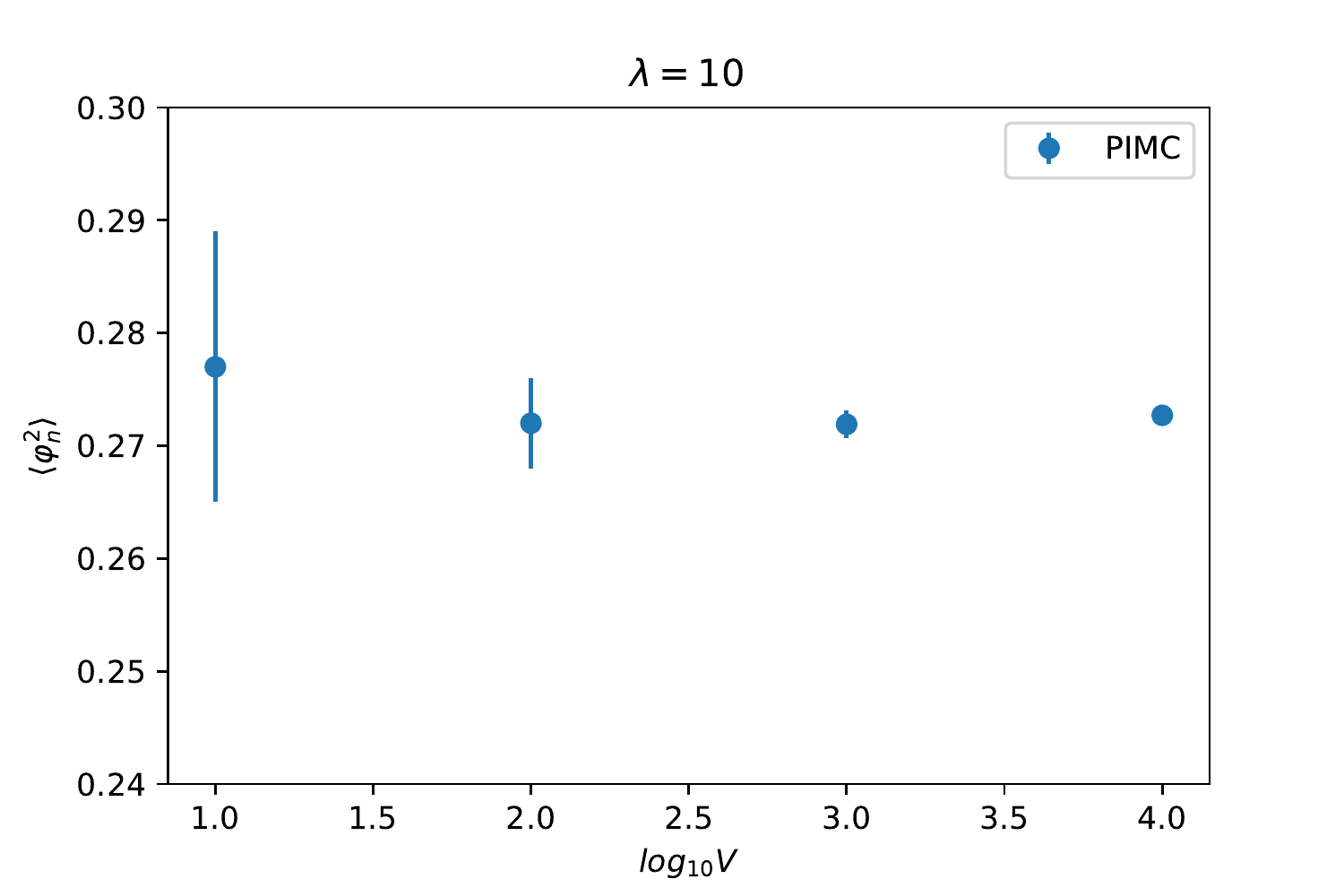}
\caption{Infinite volume limit in the PIMC computations, $D = 1$.}
\label{PIMC1}
\end{figure*}

\begin{figure*}
\centering 
\includegraphics[height=5.5cm]{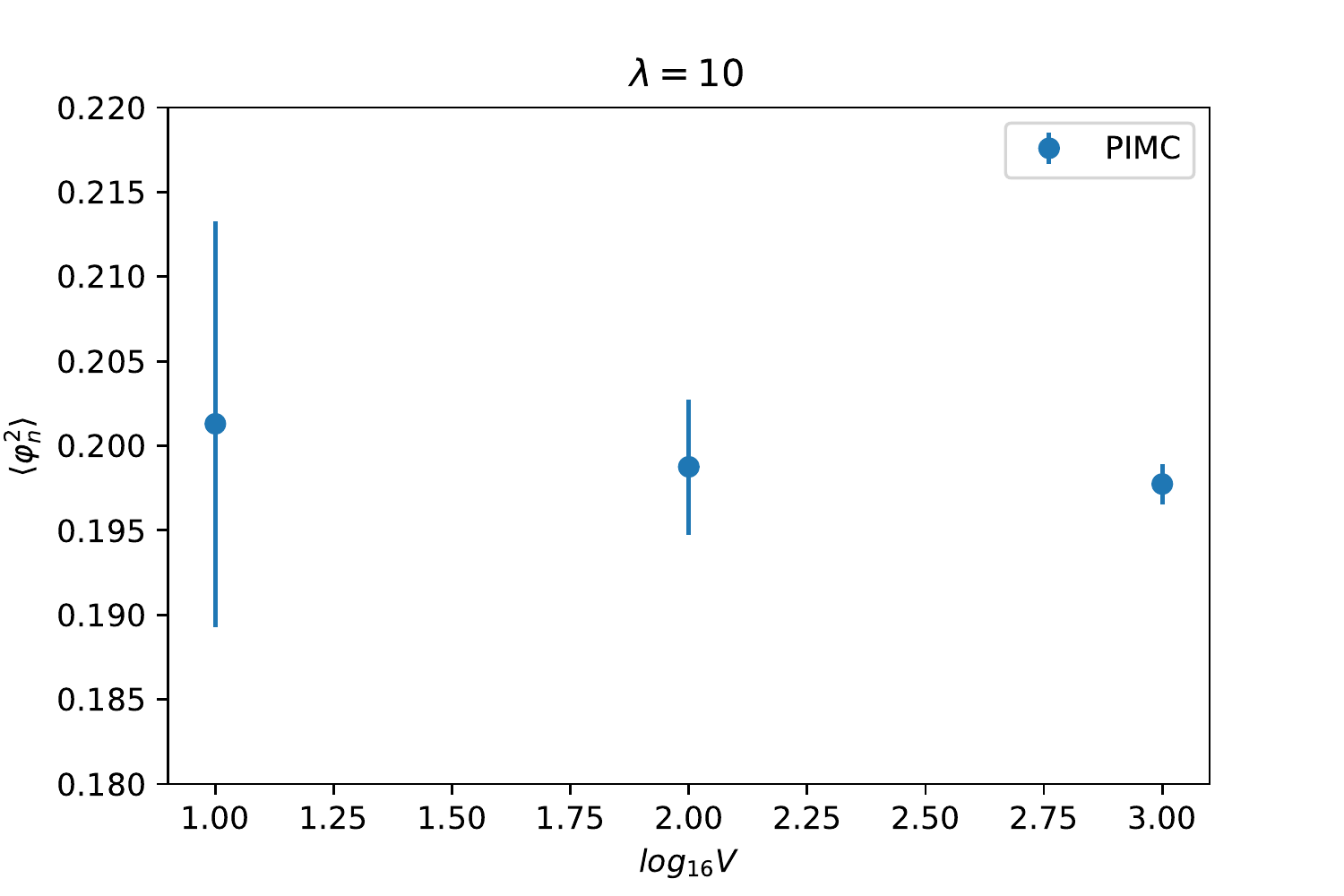}
\caption{Infinite volume limit in the PIMC computations, $D = 2$.}
\label{PIMC2}
\end{figure*}

To approach the infinite volume results using the PIMC method we analyze lattices of diferent sizes, see Fig. \ref{PIMC1} and \ref{PIMC2}.
We assume that the infinite volume limit is achieved at the lattices with the volume $V = 10000$ 
in the one dimensional case and $V = 64 \times 64$ in two dimensions.

\newpage
\label{Bibliography}

\end{document}